\documentclass[12pt]{article}
\pdfoutput=1

\usepackage{amsmath,amssymb,amsthm,mathrsfs,bbm}
\usepackage{graphicx}
\usepackage[pdftex,colorlinks=true]{hyperref}
\usepackage{cite}

\newcommand{\be}{\begin{equation}}
\newcommand{\ee}{\end{equation}}
\newcommand{\bea}{\begin{eqnarray}}
\newcommand{\eea}{\end{eqnarray}}
\numberwithin{equation}{section}

\DeclareMathOperator{\pl}{\phantom{\lambda}}

\begin{document}
\begin{flushright}
\end{flushright}
~
\vskip5mm
\begin{center} 
{\Large \bf Linear gravity from conformal symmetry}
\vskip10mm

Julian Sonner and Benjamin Withers\\
\vskip1em
Department of Theoretical Physics, University of Geneva, 24 quai Ernest-Ansermet, 1214 Gen\`eve 4, Switzerland
\vskip5mm

\tt{\{julian.sonner, benjamin.withers\}@unige.ch}

\end{center}

\vskip10mm

\begin{abstract}
We perform a unified systematic analysis of $d+1$ dimensional, spin $\ell$ representations of the isometry algebra of the maximally symmetric spacetimes AdS$_{d+1}$, $\mathbb{R}_{1,d}$ and dS$_{d+1}$.
This allows us to explicitly construct the effective low-energy bulk equations of motion obeyed by linear fields, as the eigenvalue equation for the quadratic Casimir differential operator. 
We show that the bulk description of a conformal family is given by the Fierz-Pauli system of equations.
For $\ell = 2$ this is a massive gravity theory, while for $\ell = 2$ conserved currents we obtain Einstein gravity and covariant gauge fixing conditions.
This analysis provides a direct algebraic derivation of the familiar AdS holographic dictionary at low energies, with analogous results for Minkowski and de Sitter spacetimes. 
\end{abstract}

\pagebreak
\pagestyle{plain}

\section{Introduction}
Holography, or AdS/CFT, posits that quantum gravity in asymptotically anti-de Sitter space (AdS$_{d+1}$) is dual to a conformal quantum field theory living in a space with one dimension less, a CFT$_d$ \cite{Maldacena:1997re}. The duality is often summarized by the equality of partition sums as a function of sources,
\be
Z_{\rm CFT}\left[ J_{{\cal O}_i} \right] = Z_{\rm AdS}\left[J_{{\cal O}_i}  \right]\,,
\ee 
where the left hand side is the standard partition function of the CFT with sources inserted for the operators ${\cal O}_i$, while the right-hand side is the quantum gravity partition function. Here the sources make an appearance as boundary conditions on the fields, $\phi_i$, which are dual to the operators ${\cal O}_i$ \cite{Maldacena:1997re,Witten:1998qj,Gubser:1998bc}. This dictionary applies most straightforwardly in Euclidean signature, but can also be formulated directly in the Lorentzian \cite{Balasubramanian:1998sn,Son:2002sd,Skenderis:2008dh}. Given this structure, a natural objective to is to construct gravitational bulk observables starting from only the objects and mathematical structures available in the $d$-dimensional CFT. Such a constructive approach to holography was advocated early on in \cite{Banks:1998dd,  Balasubramanian:1998de, Bena:1999jv, deHaro:2000vlm, Hamilton:2005ju, Hamilton:2006az, Hamilton:2006fh} and has been pursued from different angles over the years, for example in \cite{Heemskerk:2009pn,Heemskerk:2010ty, Heemskerk:2012np,Kabat:2012hp,Kabat:2013wga,Sarkar:2014dma}. Broadly speaking, {\it constructive AdS/CFT} starts with building up the CFT spectrum as a representation living in $d+1$ dimensions, and then using CFT dynamics to learn about bulk gravitational physics. The latter crucially involves constraining the type of CFTs that admit an (approximately) local bulk gravity description and then, within such models, the construction of bulk observables, e.g. via the specification of approximately local bulk operators. In all this it is of crucial importance to clearly delineate kinematic facts from dynamical issues, which involves answering the question of how much of the bulk dynamics is fully fixed by the underlying conformal symmetry structure.

In this paper we revisit the very first step in the constructive AdS/CFT program, addressing in a systematic and algebraic way what can be learned for (bosonic) CFT operators of arbitrary integer spin, and the bulk equations satisfied by their gravity counterparts when realizing the conformal algebra on Anti de Sitter space. At the kinematical level, we provide analogous constructions for de Sitter space (dS) and Minkowski space ($\mathbb{R}_{1,d}$). These geometries, from the point of view of the boundary field theory, are emergent and incorporate one extra (holographic) spatial direction. The de Sitter case has been invoked as the basis of a dS/CFT correspondence \cite{Strominger:2001pn}.

At the level of symmetries, Lorentzian CFTs enjoy the algebra of conformal Killing vectors of $\mathbb{R}_{1,d-1}$, the conformal algebra, $\mathfrak{so}(2, d)$. This algebra comprises dilatations and special conformal transformations in addition to the Poincar\'e translations, rotations and boosts. This algebra is also the algebra of Killing vectors of the maximally symmetric spacetime, AdS$_{d+1}$.\footnote{Throughout we will restrict to $d\geq 3$.} Building linearised fields in AdS$_{d+1}$ as representations of $\mathfrak{so}(2, d)$ makes it manifest how, at this level, such fields are simply an alternative, $d+1$ dimensional representation of the physics of a CFT$_d$.

\vskip2em
In this paper we supply an explicit construction of the linearised, low-energy equations of motion obeyed by a $d+1$ dimensional spin $\ell$ representation of the isometry algebra, $\mathfrak{so}(2,d)$,  $\mathfrak{iso}(1,d)$ and $\mathfrak{so}(1,d+1)$ and of the maximally symmetric spacetimes AdS$_{d+1}$, $\mathbb{R}_{1,d}$ and dS$_{d+1}$. We do so for all cases simultaneously utilising a continuous real parameter $\epsilon$, where $\epsilon < 0, \epsilon = 0, \epsilon > 0$ correspond to the cases AdS$_{d+1}$, $\mathbb{R}_{1,d}$ and dS$_{d+1}$ respectively. The reason we keep $\epsilon$ as a general parameter -- aside from calculational efficiency -- is so that we can utilise the natural construction of representations from the CFT point of view in AdS$_{d+1}$ for the construction in the other cases too, and hopefully shed further light on their holographic nature. In particular the results for $\epsilon=0$ may be regarded as resulting from a group contraction of the results for $\epsilon \neq 0$.

Our main output will be the two-derivative equations of motion obeyed by the linear fields, with mass terms that depend on the spin $\ell$ and a label of the representation $\Delta$. The equations are the quadratic Casimir eigenvalue equation, coupled with some additional covariant conditions which are inherited from a set of primary constraints. For instance, the equations of motion for $\ell = 1,2$ representations correspond to a massive vector and massive gravity theory, and for particular choices of $\Delta$ corresponding to conserved currents in CFT, the masses vanish and we recover Maxwell and Einstein theories respectively, while the additional covariant conditions become gauge fixing conditions.

These results have various applications, some of which we would like to briefly mention. Firstly, they give a transparent and direct derivation of the so-called holographic dictionary, which associates to every CFT operator ${\cal O}_h$ a dual bulk field $h$. These are merely two different versions of one and the same unitary irreducible representation, while the bulk wave equation, satisfied by $h$ is simply the action of the quadratic Casimir in one of these two representations. 

Secondly,  as alluded to above, these results can be seen as the first step in the program to reconstruct bulk physics purely from CFT objects. Linear equations on AdS$_{d+1}$, i.e. fluctuations around the conformal vacuum, including linearised Einstein gravity, are purely kinematical consequences of the underlying symmetry principle. 

Of course the ideas underpinning this work are not new, based as they are, on a Weinberg-like approach \cite{Weinberg:1995mt}. These and related ideas have been discussed in the context of AdS holography by many authors, for example \cite{Banks:1998dd, Balasubramanian:1998de, Ferrara:1998jm, Bena:1999jv, Balasubramanian:1999ri, Hamilton:2005ju, Hamilton:2006az, Hamilton:2006fh, Fitzpatrick:2010zm, ElShowk:2011ag, Sundrum:2011ic, Fitzpatrick:2012cg, Fitzpatrick:2014vua, Nakayama:2015mva, Kitaev:2017hnr}, higher-spin holography \cite{Metsaev:2003cu, Rahman:2015pzl}, and explored in detail for the case $\ell = 0$ in AdS \cite{PhysRevD.33.389, Maldacena:1998bw, KaplanLectures, Sundrum:2011ic}, and for the de Sitter case \cite{Dirac1963, PhysRevD.20.848}. In particular we would like to highlight the pedagogical presentation of \cite{KaplanLectures}. Rather, the goal of this paper is to provide a general, systematic analysis which encompasses not only scalars but fields of any integer spin and any Lorentzian maximally symmetric background to boot. 
This analysis finds that the natural entries in the holographic dictionary are bulk fields governed by the natural generalisation of the Fierz-Pauli system of equations \cite{Fierz:1939zz,Fierz:1939ix} with specific mass and spin as dictated by the labels of the corresponding field theory representation.

\subsection{Summary of results}
A natural starting point for the analysis in this paper are $d-$dimensional CFTs, and in particular their basic building blocks, namely conformal families.
Let us therefore consider an irreducible representation constructed from the lowest-weight state $ | \Delta, \ell, \cdots \rangle$, satisfying the {\it primary conditions}
\bea
K_a | \Delta, \ell, \cdots \rangle &=&0 \nonumber\\
 D | \Delta, \ell, \cdots \rangle  &=& i \Delta | \Delta, \ell, \cdots \rangle \nonumber\\
 \tfrac{1}{2}J^2  | \Delta, \ell, \cdots \rangle  &=& \ell (\ell + d - 2)  | \Delta, \ell, \cdots \rangle\,,\label{primaryconds}
\eea
where for $\epsilon < 0$, the generators $K_a, D$, $J_{ab}$ \& $P_a$ obey the usual conformal algebra $\mathfrak{so}(2,d)$, while for general $\epsilon$ the precise algebra will be given below in Eqs. \eqref{eq.ConformalAlgebraFirst}-\eqref{eq.ConformalAlgebraLast}.
$\Delta$ is the conformal dimension, $\ell$ is the spin, and the ellipses denote dimension dependent additional labels, such as angular momentum components or charges, which will not be needed for this analysis.
As usual, descendants are constructed by acting repeatedly with $P_a$.

Then we construct a second representation living on one of AdS$_{d+1}$ ($\epsilon <0$), ${\mathbb R}_{1,d}$  ($\epsilon =0$), or dS$_{d+1}$ ($\epsilon >0$), in terms of a symmetric rank$-\ell$ tensor, $h^{\mu_1 \cdots \mu_\ell}$, under the action of Lie derivatives. We establish algebraically that 
\bea
\nabla_\lambda h^{\lambda \mu_2 \cdots \mu_\ell} &=& 0\,,\qquad \qquad (\ell \geq 1)\nonumber\\
h_\lambda{}^{\lambda \mu_3 \cdots \mu_\ell} &=& 0\,,\qquad \qquad (\ell \geq 2) \nonumber\\
\left(\Box + \frac{\epsilon}{L^2}((\ell-2)(\ell + d -3)-2) \right)h^{\mu_1\ldots \mu_\ell} &=& m^2h^{\mu_1\ldots \mu_\ell}\,.\label{mcas}
\eea
The trace and covariant derivatives are taken with respect to the metric on AdS$_{d+1}$,  ${\mathbb R}_{1,d}$, or dS$_{d+1}$ respectively. The first two equations are covariant constraints implied by the primary conditions, but also apply to all descendants, i.e. they commute with the application of $P_a$. The final equation, taking the form of the linear massive spin-$\ell$ wave equation,  is the eigenvalue equation for the quadratic Casimir operator, where\footnote{For $\ell=0$ the mass-squared is not given by the parameter $m^2$ but combining two terms in \eqref{mcas}, specifically, $-\epsilon L^{-2}\Delta(\Delta-d)$. For more details see the discussion following Eq. \eqref{eq.massParameter} below.}
\be
m^2L^2 = -\epsilon(\Delta + \ell -2)(\Delta - d - \ell +2) . \label{masssqparam}
\ee
In other words, every bosonic spin-$\ell$ irreducible representation of the algebra for the appropriate value of $\epsilon$, living in $d$ dimensions, directly gives rise to a linear (generically) massive wave equation in $d+1$ dimensions, with properties as specified above. This identifies the bulk equation appropriate to the conformal family associated to a primary operator in a CFT.
\newline\newline
We begin by introducing our $\epsilon$-parameterised metric for all three maximally symmetric spacetimes in section \ref{sec:metric}, followed immediately by their complete set of linearly independent Killing vectors and associated Lie bracket algebra in section \ref{sec:Killing}. We put this information to work in section \ref{sec:rep} where we construct families of states labelled by $(\Delta, \ell, \cdots)$ and directly compute the second-order equations of motion they obey through the quadratic Casimir of the isometry algebra. We conclude with a discussion in section \ref{discussion}.

\section{Constructing bulk representations}
\begin{center}
\begin{figure*}[t] 
\includegraphics[width=\textwidth]{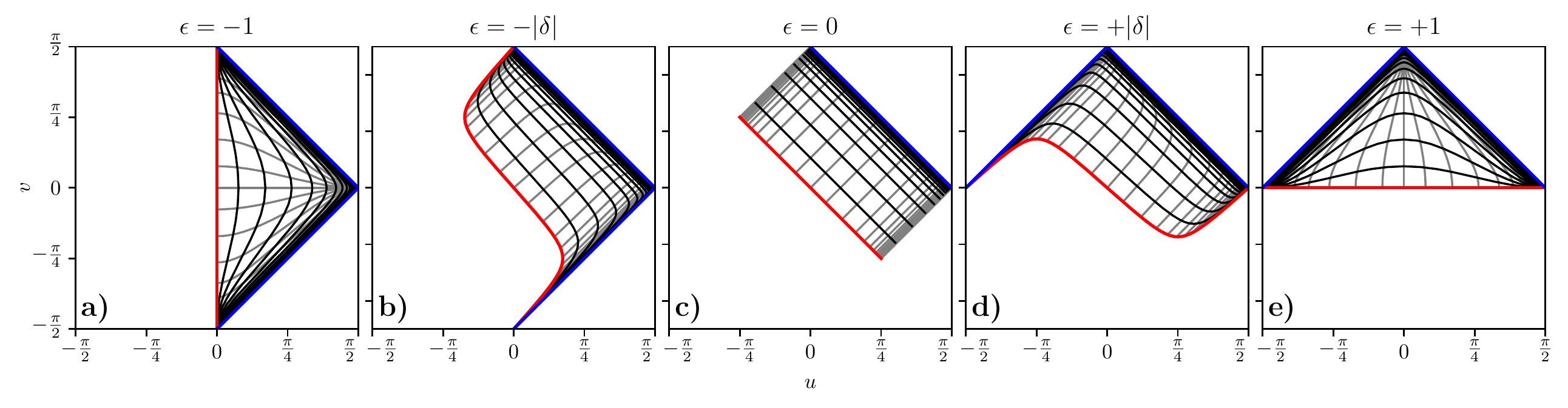}
\caption{Region of spacetime covered by the unified flat slicing \eqref{met}, compactified to the $(u,v)$ plane, in the case of \textbf{a)} AdS$_{d+1}$, \textbf{b)} AdS$_{d+1}$, \textbf{c)} $\mathbb{R}_{1,d}$, \textbf{d)} dS$_{d+1}$, \textbf{e)} dS$_{d+1}$ with both the coordinate slicing and radius of curvature dictated by the value of $\epsilon$ indicated. Lines of constant $y^1$ and $y^0$ are shown in grey and black respectively, here restricted to $y^0\geq 0$. Red denotes the set of points reached by $y^0\to 0^+$ and blue the set of points as $y^0\to +\infty$. Note that the $y^0=0$ set is everywhere timelike for $\epsilon<0$, and everywhere spacelike for $\epsilon>0$, sandwiching the lightlike case at $\epsilon = 0$. Details of construction are given in appendix \ref{app:theta}.\label{fig.slicings}}
\end{figure*}
\end{center}
\subsection{A unified flat slicing\label{sec:metric}}
Let us begin by writing the unified maximally symmetric spacetime considered in this work in flat-sliced or Poincar\'e coordinates as the following line element (see Fig. \ref{fig.slicings}):
\be
ds^2 = \frac{L^2}{(b\cdot x)^2}\eta_{\mu\nu}dx^\mu dx^\nu \label{met},
\ee
where the indices run from $\mu,\nu = 0,\ldots, d$, and we have introduced the constant metric $\eta = \text{diag}(-1,1,\ldots)$ which we use to construct `$\cdot$' contractions and also squares of vectors in this section only. In addition we have introduced the constant vector $b$ whose norm introduces the parameter $\epsilon$ via $b^2 = -\epsilon$. For convenience we shall take the vector $b$ to lie in the $(x^1, x^0)$ plane, and introduce a second constant vector $k$ in the same plane obeying $k\cdot k = \epsilon$ and $k \cdot b = \sqrt{1-\epsilon^2}$. Further details of this construction are given in appendix \ref{app:theta}.

Taking for illustration the value $\epsilon = -1$, the metric \eqref{met} is AdS$_{d+1}$ in Poincar\'e coordinates: since $b$ is then a spacelike vector and the projection $b\cdot x$ is a spatial coordinate parameterising the distance from the timelike conformal boundary of AdS. Similarly for $\epsilon = +1$, the line element \eqref{met} is that of dS$_{d+1}$ in flat-sliced coordinates: $b$ is now a timelike vector and the projection $b\cdot x$ is the usual conformal time parameter, taking the value zero on the spacelike conformal boundary in the future, or the past. For $\epsilon = 0$, the line element \eqref{met} is simply that of $\mathbb{R}_{1,d}$ written in a perhaps unfamiliar form: to reach \eqref{met} starting from the familiar line element $\eta_{\mu\nu} dx^\mu dx^\nu$ we first perform a special conformal transformation, namely
\be
x^\mu\to \frac{x^\mu-\tilde{b}^\mu x^2}{1-2\tilde{b}\cdot x + \tilde{b}^2 x^2}
\ee
where the parameter $\tilde{b}^\mu = \frac{-b^\mu}{2L}$. Following this we perform a translation $x^\mu \to x^\mu - L k^\mu$, which gives \eqref{met}. 
More generally, the spacetime \eqref{met} is AdS$_{d+1}$ for $-1\leq \epsilon < 0$, $\mathbb{R}_{1,d}$ for $\epsilon = 0$ and dS$_{d+1}$ for $0 < \epsilon \leq1$, with a radius given by $L/\sqrt{|\epsilon|}$. 
 
In what follows it will be convenient to distinguish the `holographic' radial direction from the others. To do this we introduce a new set of coordinates $y^\mu$, defined as
\be
y^0 = \eta_{\mu\nu}b^\mu x^\nu, \quad y^1 = \eta_{\mu\nu}k^\mu x^\nu,\quad y^i = x^i. \label{ymap}
\ee
where $\mu = 0,\ldots, d$, and $i = 2,\ldots d$.
Now $y^0$ parameterises the distance from the conformal boundary, and the remaining $d$ coordinates $y^a$ with $a = 1 \ldots d$ label points on the boundary. 
Written in these coordinates the line element \eqref{met} becomes
\bea
\frac{L^2}{(\eta_{\rho\sigma}b^\rho x^\sigma)^2}\eta_{\mu\nu}dx^\mu dx^\nu &=& g_{\mu\nu}dy^\mu dy^\nu\nonumber\\
&=& \frac{L^2}{(y^0)^2}\left(-\epsilon (dy^0)^2 +2\sqrt{1-\epsilon^2} dy^0 dy^1  + \epsilon (dy^1)^2 + \delta_{ij} dy^i dy^j\right)\nonumber\\&&\label{gdef}
\eea
This slicing, parameterised by $\epsilon$, allows us to rotate from one case to the other by the introduction of off-diagonal $dy^0 dy^1$ metric components. The utility of this combined parameterisation and choice of slicing is illustrated in the plots of Fig. \ref{fig.slicings}, whose precise construction is detailed in appendix \ref{app:theta}. Taking $\epsilon\rightarrow 0$ from either side amounts to taking the flat space limit of (anti) de Sitter space, and we illustrate the fate of the UV boundary (in red) and the Poincar\'e horizon (in blue) under this limit, for both signs of $\epsilon$. It will not escape the reader's attention that this scaling is equivalent to dialling the curvature, with the flat-space limit naturally being $L/\sqrt{|\epsilon|} \rightarrow \infty$.

From this point on, all dot products, raising and lowering of Greek indices will be with respect to the metric $g_{\mu\nu}$ as defined in \eqref{gdef}. For later use, we note also that the Riemann curvature of $g$ is given by,
\be
R_{\mu\nu\rho}^{\phantom{\mu\nu\rho}\sigma} = \frac{\epsilon}{L^2}\left(g_{\mu\rho}\delta_\nu^\sigma - g_{\nu\rho}\delta_\mu^\sigma\right).\label{riemann}
\ee

\subsection{Killing vectors and Lie bracket algebra\label{sec:Killing}}
We will now detail the Killing vectors of \eqref{met} vectors in full. As a maximally symmetric spacetime for any $\epsilon$, \eqref{met} has $ \frac{1}{2}(d+1)(d+2)$ linearly independent Killing vectors. In particular we wish to write them in a way which inherits the language of the conformal algebra, and the $y^\mu$ coordinates introduced above serves this purpose.
In these coordinates we construct the full set of Killing vectors. They are most straightforwardly obtained from an embedding formalism, which we detail in appendix \ref{app:embed}. Using the $a,b = 1, \ldots d$ as labels we have,
\bea
D  &=& -i y\cdot \partial\nonumber\\
P_a &=& -i\partial_a\nonumber\\
J_{ab} &=& -i\frac{(y^0)^2}{L^2}\left(y_a\partial_b - y_b\partial_a \right)\nonumber\\
K_a &=& -i\frac{(y^0)^2}{L^2}(y\cdot y \partial_a - 2 y_a y\cdot \partial)\label{Killing}
\eea
where $y_\mu \equiv g_{\mu\nu} y^\nu$ and $\partial_\mu \equiv \partial/\partial y^\mu$.
One can verify that there are $1+ d+\frac{d}{2}(d-1) + d = \frac{1}{2}(d+1)(d+2)$ of them, and we have the complete set.
These vectors satisfy the following algebra, governed by the constant $d$-metric $\gamma_{ab} = \frac{(y^0)^2}{L^2}g_{ab} = \text{diag}(\epsilon, 1,1,\ldots)_{ab}$,
\bea\label{eq.ConformalAlgebraFirst}
\left[J_{ab},J_{ce}\right] &=& i \left(\gamma_{ac}J_{be} -\gamma_{ae}J_{bc}+\gamma_{be}J_{ac}-\gamma_{bc}J_{ae}\right)\\
\left[J_{ab}, P_c\right] &=& i \left(\gamma_{ca} P_b - \gamma_{cb} P_a \right) \\
\left[D,P_a\right] &=& i P_a\\
\left[J_{ab}, K_c\right] &=& i \left(\gamma_{ca} K_b - \gamma_{cb} K_a \right) \\
\left[P_a, K_b\right] &=& 2i\left(\gamma_{ab} D - J_{ab}\right)\\
\left[D, K_a\right] &=& -i K_a \label{eq.ConformalAlgebraLast}
\eea
where the other combinations are zero and $\left[\cdot,\cdot\right]$ is the Lie bracket, which in the case above corresponds to the vector commutator. For $\epsilon < 0$ this is the conformal algebra, $\mathfrak{so}(2,d)$. For $\epsilon > 0$ this is the conformal algebra in Euclidean signature, $\mathfrak{so}(1,d+1)$. For $\epsilon = 0$ this is the Poincar\'e algebra in $d+1$ dimensions, $\mathfrak{iso}(1,d)$. Note that the `boundary' metric $\gamma_{ab}$ becomes degenerate at $\epsilon = 0$, thus some of the structure constants in the algebra vanish. We may therefore regard the Minkowski case as being reached by an \.In\"on\"u-Wigner contraction from either $\mathfrak{so}(2,d)$ or $\mathfrak{so}(1,d+1)$. We provide a map to a more familiar presentation of the Poincar\'e algebra in appendix \ref{app:mink}. 

The quadratic Casimir operator of the above algebra is given by the following expression
\be
C_2 = -{\cal L}_D{\cal L}_D - \frac{\gamma^{ab}}{2}\left({\cal L}_{P_a}{\cal L}_{K_b}+ {\cal L}_{K_a}{\cal L}_{P_b}\right) + \frac{\gamma^{ab}\gamma^{ce}}{2} {\cal L}_{J_{ac}}{\cal L}_{J_{be}} \label{cas},
\ee
where ${\cal L}$ denotes the Lie derivative, and it obeys the Casimir property $[{\cal L}_\tau, C_2]=0$ for all generators $\tau$. Note that the spatial generator labels denoted by $a,b$ have been contracted with $\gamma^{ab}$, the inverse of $\gamma_{ab}$ defined above.\footnote{In particular note that $\gamma^{ab} = \text{diag}(\frac{1}{\epsilon}, 1,1,\ldots)^{ab} \neq \frac{L^2}{(y^0)^2}g^{ab}$.} \eqref{cas} is a second order differential operator, and one can verify by direct computation that when acting on a symmetric rank $\ell$ tensor $h$ it gives,
\be
(C_2h) ^{\mu_1\ldots \mu_\ell} = -\frac{L^2}{\epsilon} \Box h^{\mu_1\ldots \mu_\ell}  + \ell (\ell + d - 1)h^{\mu_1\ldots \mu_\ell} - \ell(\ell-1) h_\lambda^{\pl\lambda(\mu_3\ldots \mu_\ell}g^{\mu_1\mu_2)}, \label{cash}
\ee
where the parentheses among the indices denote symmetrisation with unit strength. In the Minkowski case, $\epsilon = 0$, this quadratic Casimir becomes the familiar momentum-squared differential operator, as demonstrated in appendix \ref{app:mink}.

\subsection{Primary conditions\label{sec:rep}}
One representation of the algebra of section \ref{sec:Killing} is in terms of the Lie derivatives of a symmetric tensor of rank $\ell$, denoted here $h^{\mu_1\ldots \mu_\ell}$. In the usual way $K_a$ and $P_a$ act as lowering and raising operators with respect to dilatations $D$, and so it is natural to arrange the representations in  families accordingly. Such a family can be labelled by the dilatation eigenvalue of its primary, $i\Delta$, where the primary corresponds to the field configuration $\bar{h}$\footnote{Throughout this paper we will suppress spacetime indices where suitable in order to improve legibility.}  which is annihilated by $K_a$ for all $a$. The remaining members of the family are descendants reached by successive applications of $P_a$. Thus we begin with a set of primary conditions that define $\bar{h}$,
\bea
\left[D,\bar{h}\right] &=& i \Delta \bar{h},\label{dil}\\
\left[K_a,\bar{h}\right] &=& 0,\label{kill}
\eea
in addition, we require that $\bar{h}$ have spin $\ell$ and we do so through
\be
\frac{1}{2}\gamma^{ab}\gamma^{cd}\left[J_{ac},\left[J_{bd},\bar{h}\right]\right] = \ell(\ell+d-2)\bar{h}.\label{Jsq}
\ee
Note that the first two conditions \eqref{dil}, \eqref{kill} are first order PDEs in $d+1$ dimensions, whilst \eqref{Jsq} is a second order PDE. 
To proceed one could solve the PDEs \eqref{dil}, \eqref{kill}, \eqref{Jsq} directly to construct $\bar{h}$, and subsequently all descendants through differentiation by $y^a$. The integration constants that arise in this procedure go hand in hand with the different available polarisations for spin $\ell$ in $d+1$ dimensions. However, we can directly obtain the covariant equations of motion obeyed by all members of the family through purely algebraic means, and so we proceed in this manner instead.

Our goal is to obtain covariant equations of motion. Note however, that the primary conditions as presented above are not covariant conditions; for example the generators, through their labelling, explicitly distinguish $y^0$ from the $y^a$. 
In light of this we find it useful assemble the corresponding Lie brackets into the object ${\cal T}_\mu$ as follows,
\bea
{\cal T}_0 &=& -\frac{L^2}{(y^0)^2}\frac{i}{y\cdot y} \frac{y^{a}}{y^0}\left[K_{a},h\right] + i\left( \frac{2y_0}{y\cdot y}-\frac{1}{y^0}\right)\left(\left[D,h\right]- i \Delta h\right)\nonumber\\
{\cal T}_a &=& \frac{L^2}{(y^0)^2}\frac{i}{y\cdot y} \left[K_{a},h\right] + 2i\frac{y_{a}}{y\cdot y}\left(\left[D,h\right]- i \Delta h\right),
\eea
whose definition is valid for any $h$ but holds the particular property that ${\cal T}_\mu = 0$ when evaluated on the primary, $\bar{h}$. Then, one arrives at a compact identity for the covariant derivative of the bulk field,
\be
\nabla_\lambda h^{\mu_1 \ldots \mu_\ell} = \left({\cal T}_\lambda\right)^{\mu_1\ldots\mu_\ell} - \Delta \zeta_\lambda h^{\mu_1\ldots \mu_\ell}  - \ell  \zeta^{(\mu_1} h_\lambda^{\pl\mu_2\ldots\mu_\ell)} + \ell \zeta_\sigma h^{\sigma (\mu_2\ldots\mu_\ell} \delta^{\mu_1)}_\lambda.\label{tensor1derivs}
\ee
where we have introduced for brevity,
\be
\zeta_\rho \equiv \frac{2y_\rho}{y\cdot y} - \frac{\delta^0_\rho}{y^0},
\ee
which has norm $\zeta\cdot \zeta = -\epsilon/L^2$.
Evaluated on the primary $\bar{h}$ we have ${\cal T}_\mu = 0$ and the relations \eqref{tensor1derivs} are merely a convenient rewriting of the dilatation and special conformal PDEs \eqref{dil} and \eqref{kill}, which form part of the definition of the primary. For later use let us here give the result of computing the divergence using \eqref{tensor1derivs},
 \be
\nabla_\lambda h^{\lambda \mu_2\ldots \mu_\ell} = \left({\cal T}_\lambda\right)^{\lambda \mu_2\ldots \mu_\ell} + (d+\ell -1 -\Delta) \zeta_\lambda h^{\lambda \mu_2 \ldots \mu_{\ell}} - (\ell - 1) h_{\lambda}^{\pl\lambda(\mu_3\ldots \mu_{\ell}}\zeta^{\mu_2)}.\label{div}
\ee
To understand the consequences of the remaining primary condition, \eqref{Jsq}, it is most convenient to turn to the Casimir operator \eqref{cas}. In particular, it is straightforward to show by use of the commutation relations that all three conditions \eqref{dil}, \eqref{kill}, \eqref{Jsq} ensure that any member of the family $h$ are eigenfunctions of the $C_2$ operator with eigenvalues $\lambda_2$,
\be
C_2 h = \lambda_2 h,\qquad \lambda_2 = \Delta(\Delta - d) + \ell(\ell+d-2).\label{caseval}
\ee
We can then use this expression to extract any additional constraints that arise from \eqref{Jsq}; we  differentiate \eqref{tensor1derivs} once and insert into \eqref{caseval}, revealing a set of linear algebraic constraints,
\bea
\ell(\ell-1)\left( \bar{h}_\lambda^{\pl\lambda(\mu_3\ldots \mu_\ell}g^{\mu_1\mu_2)} +\frac{L^2}{\epsilon} \bar{h}_\lambda^{\pl\lambda(\mu_3\ldots \mu_\ell}\zeta^{\mu_1}\zeta^{\mu_2)}+ \frac{L^2}{\epsilon} \zeta_\lambda \zeta_\sigma \bar{h}^{\lambda\sigma(\mu_3\ldots \mu_\ell}g^{\mu_1\mu_2)}\right)=\nonumber\\
\frac{L^2}{\epsilon} \ell(d+2\ell -3) \zeta_\sigma \bar{h}^{\sigma (\mu_2\ldots \mu_\ell}\zeta^{\mu_1)}. \label{constraints}
\eea
Note that terms with index structure $\bar{h}^{\mu_1\ldots \mu_\ell}$ do not appear and the constraints are independent of $\Delta$. Taking stock of our current position, once the conditions \eqref{constraints} and \eqref{tensor1derivs} evaluated on $\bar{h}$ are met, we have satisfied the primary PDE conditions \eqref{dil}, \eqref{kill} and \eqref{Jsq}. The remainder of this section will be devoted to understanding the $\zeta$-independent, covariant conditions that apply to $\bar{h}$, and subsequently to a general member of the conformal family $h$. The resultant equations of motion that the $h$ satisfy will be discussed in the next section.

Let us first consider the cases $\ell = 0,1,2$ individually. For $\ell = 0$ there is no constraint coming from \eqref{constraints}. For $\ell = 1$ one can see immediately from \eqref{constraints} that $\zeta_\lambda \bar{h}^\lambda = 0$ and hence by \eqref{div} one obtains $\nabla_\lambda \bar{h}^\lambda = 0$. For $\ell = 2$ the procedure is simple also; contracting \eqref{constraints} with $\zeta_{\mu_1}\zeta_{\mu_2}$ reveals $\zeta_\lambda \zeta_\sigma \bar{h}^{\lambda\sigma} = 0$, subsequently contracting \eqref{constraints} with $\zeta_{\mu_1}$ gives $\zeta_\lambda \bar{h}^{\lambda\mu_1} = 0$, then \eqref{constraints} gives $\bar{h}_\lambda^{\pl\lambda} = 0$ and finally \eqref{div} gives $\nabla_\lambda \bar{h}^{\lambda\mu_1} = 0$.

For the higher spin cases, $\ell\geq 3$, we simply note that the constraints \eqref{constraints} are satisfied when $\zeta_\lambda \bar{h}^{\lambda\mu_2\ldots\mu_\ell} = 0$ and $\bar{h}_{\lambda}^{\pl\lambda\mu_3\ldots \mu_\ell}=0$, from which follows $\nabla_\lambda \bar{h}^{\lambda\mu_2\ldots\mu_\ell} = 0$. For these cases one can show that $\nabla_\lambda \bar{h}^{\lambda\mu_2\ldots\mu_\ell} = 0$ is implied by \eqref{constraints} if we first assume that $\bar{h}_{\lambda}^{\pl\lambda\mu_3\ldots \mu_\ell}=0$, but the general analysis is somewhat unwieldy and we leave it for future work. Thus, unlike the cases $\ell = 0,1,2$ above, we leave open the possibility that \eqref{constraints} are satisfied by weaker covariant constraints on $h$, but we shall proceed under the assumption that this is not the case.

Finally, we show that the conditions on the primaries $\bar{h}$ are inherited by all descendants, and thus apply to any $h$. Given any member of the family $h^{\mu_1\ldots \mu_\ell}$ we can construct a descendant $\tilde{h}^{\mu_1\ldots \mu_\ell} \equiv \left[P_a, h\right]^{\mu_1\ldots \mu_\ell}$ for some label $a$. The descendant $\tilde{h}$ will have different angular momentum labels as compared to $h$, amounting to the choice of $a$, a label which has been absorbed into the definition of $\tilde{h}$. Crucially we note that $g_{\mu_1\mu_2}$ only depends on $y^0$, and from this alone it follows that taking $y^a$ derivatives of $h$ commutes with the trace and divergence operations. In other words,
\bea
g_{\mu_1\mu_2}h^{\mu_1\mu_2\mu_3\ldots \mu_\ell} = 0 &\implies& g_{\mu_1\mu_2}\tilde{h}^{\mu_1\mu_2\mu_3\ldots \mu_\ell} = 0,\\
\nabla_\lambda h^{\lambda\mu_2\ldots \mu_\ell} = 0 &\implies& \nabla_\lambda \tilde{h}^{\lambda\mu_2\ldots \mu_\ell} = 0.
\eea
Thus the tracelessness and divergence-free covariant conditions that apply to the primaries apply to all their descendants too. 

\subsection{Two derivative equations of motion}
In section \ref{sec:rep} we explored the consequences of the primary conditions \eqref{dil}, \eqref{kill} and \eqref{Jsq} as constraints acting on both the primaries and all descendants, $h$, arriving at the covariant constraints\footnote{As a reminder, we showed that these constraints hold for the cases $\ell = 0,1,2$. For $\ell \geq 3$ we showed that they are true assuming a traceless primary $\bar{h}$.}
\bea
\nabla_\lambda h^{\lambda\mu_2\ldots \mu_\ell} &=& 0 \qquad \ell \geq 1, \label{constrace}\\
h_\lambda^{\pl\lambda\mu_3\ldots \mu_\ell} &=& 0\qquad \ell \geq 2.\label{consdiv}
\eea
where the trace and covariant derivative are taken with respect to the background metric $g$, \eqref{gdef}.
Imposing these conditions on quadratic Casimir differential equation \eqref{caseval} gives the following two-derivative equation of motion obeyed by all $h$ in the family,
\be
\left(\Box + \frac{\epsilon}{L^2}((\ell-2)(\ell + d -3)-2) \right)h^{\mu_1\ldots \mu_\ell} = m^2h^{\mu_1\ldots \mu_\ell},\label{eoms}
\ee
with
\be\label{eq.massParameter}
m^2L^2 = -\epsilon(\Delta + \ell -2)(\Delta - d - \ell +2) .
\ee
Where $m$ is the mass of the field for $\ell \geq 1$; for $\ell = 0$ the mass term is given by combining the right hand side with the second term on the left hand side giving the familiar mass-squared value: $-\epsilon L^{-2}\Delta(\Delta -d)$. Here it will be convenient to express \eqref{eoms} in a slightly different form, obtained by using \eqref{consdiv} and the commutation relation for covariant derivatives and \eqref{riemann} we obtain,
\be
\Box h^{\mu_1\ldots \mu_\ell} - \ell \nabla_\lambda \nabla^{(\mu_1} h^{\mu_2\ldots\mu_\ell)\lambda}+ \frac{\epsilon}{L^2}2(\ell-1)(\ell + d -2)h^{\mu_1\ldots \mu_\ell} = m^2h^{\mu_1\ldots \mu_\ell}. \label{eomsrewrite}
\ee 

The mass vanishes at $\Delta = \bar{\Delta} \equiv d + \ell - 2$, corresponding to the conformal dimension of a conserved CFT current operator in $d$ spacetime dimensions. The other massless case sits at $\Delta = d-\bar{\Delta}$ corresponding to an alternate quantisation. The $m=0$ case is characterised by a gauge invariance, though note that the conditions \eqref{constrace} and \eqref{consdiv} partially fix this gauge. In detail, one can show that the equations \eqref{eomsrewrite} and conditions \eqref{constrace}, \eqref{consdiv} at $m=0$ are invariant under
\be
h^{\mu_1\ldots \mu_\ell} \to h^{\mu_1\ldots \mu_\ell} + \nabla^{(\mu_1}\phi^{\mu_2\ldots \mu_\ell)}
\ee
provided $\phi$ itself is transverse, traceless and obeys the equation $\nabla_{\mu_1} \nabla^{(\mu_1}\phi^{\mu_2\ldots\mu_\ell)} = 0$.

Let us now consider the individual cases $\ell = 0,1,2$ in some more detail. For $\ell = 0$ the equations of motion for $h$, \eqref{eoms}, correspond to is a massive Klein-Gordon field for general $\Delta$, with a mass $-\epsilon L^{-2}\Delta(\Delta -d)$. It is not subject to any additional constraints.
For $\ell = 1$ one has a massive vector theory in general, subject to the Lorenz condition \eqref{consdiv}. For values of $\Delta$ corresponding to $m=0$, i.e. the dual of a $d$-dimensional conserved current operator $J^a$, these are precisely the Maxwell equations on the background \eqref{gdef} in Lorenz gauge, as can be most directly seen in \eqref{eomsrewrite}.
For $\ell = 2$ this is a massive gravity theory in general subject to both conditions \eqref{constrace} and \eqref{consdiv}. For massless $\Delta$, corresponding to the $d$-dimensional stress tensor operator $T^{ab}$, these equations correspond to linearised Einstein gravity on the background \eqref{gdef}, in a covariant transverse and traceless gauge. In some more detail, to compare with the Einstein equations consider the equations of motion for the action, 
\be
S = \frac{1}{2\kappa^2}\int d^{d+1}x \sqrt{-G} \left(R - 2\Lambda\right),\qquad \Lambda \equiv \frac{\epsilon}{L^2} \frac{d(d-1)}{2},
\ee
where $R$ is the Ricci tensor of the metric $G$, and linearise with $h$, i.e. take the metric to be $G = g + h$ where $g$ is the metric \eqref{gdef} and $h$ is a perturbation. It is straightforward to see from \eqref{riemann} that the maximally symmetric background $g$ solves the unperturbed equations of motion, i.e. the Ricci tensor of $g$,  $R(g) = \epsilon d g/L^2$. Then, to linear order in $h$ the equations of motion are given precisely by \eqref{eoms} together with the conditions \eqref{constrace} and \eqref{consdiv} in the case of massless operator dimensions: $\Delta = \bar{\Delta}$ or $\Delta = d-\bar{\Delta}$.

\section{Discussion\label{discussion}}
The symmetries of a CFT$_d$ admit a geometric representation in $d+1$ dimensions. Viewed as basic starting point for holographic duality, one may identify $d+1$ dimensional, symmetric rank $\ell$ tensors $h$ transforming under the isometry algebra of AdS$_{d+1}$ with conformal families of operators in the CFT$_d$. In this paper we constructed such a representation explicitly, through the action of Lie derivatives of $h$ by the bulk Killing vectors. The field $h$ obeys a Casimir eigenvalue equation, which takes the form of a two-derivative differential operator. Thus in a low-energy effective theory, the Casimir equation serves as the linearised equations of motion when supplemented by appropriate constraints, which we detailed here. These constraints amount to a partial gauge fixing in the case of vanishing bulk mass, and otherwise serve as constraints to the massive case. Through a parameter $\epsilon$ we simultaneously obtained analogous results for other maximally symmetric spacetimes $\mathbb{R}_{1,d}$ and $dS_{d+1}$.

Note that we couched our calculations in flat-sliced coordinates, however the final equations of motion \eqref{eoms}, \eqref{constrace}, \eqref{consdiv} are covariant in the background metric $g$. Thus the equations of motion are valid for any coordinate system of interest, for instance global or static patch coordinates. Strictly speaking these covariant equations apply around the vacuum state of the theory, dual to the three maximally symmetric spacetimes. However, since experience from AdS/CFT suggests that the bulk equations linearise asymptotically also for more general states, one may hope that these equations provide a useful starting point for building up more general states, for example as coherent excitations of the vacuum \cite{Botta-Cantcheff:2015sav, Marolf:2017kvq, Belin:2018fxe}.

Covariant bulk wave equations are an important ingredient in the construction of bulk operators from boundary ones, along the lines of \cite{Hamilton:2006az}, whose construction, at leading order in $1/N$, involves smearing the boundary operator against a kernel which solves precisely the wave equations derived here. Our analysis, at leading order in $1/N$, demonstrates in a transparent way how this structure follows only from CFT$_d$ arguments, with analogous considerations applying for the other two maximally symmetric spaces $\mathbb{R}_{1,d}$ and dS$_{d+1}$.

Beyond the duality identification of bulk field representations with CFT operator representations we did not stray from group theoretic arguments. As such there are several physical considerations which lie outside the scope of this paper, including those related to the introduction of consistent interactions, the requirements on unitarity and whether or not there is a sharp notion of locality \cite{Hamilton:2005ju,Heemskerk:2009pn,Fitzpatrick:2010zm,ElShowk:2011ag,Fitzpatrick:2012cg}, not to mention the reconstruction of the bulk spacetime inside the horizon of a black hole  \cite{Hamilton:2006fh,Papadodimas:2012aq}. It will be important to incorporate consistent interactions, as has been explored for the case of massless higher-spins in AdS, for example \cite{Bekaert:2014cea, Sleight:2016dba}.

More speculatively, it is certainly intriguing to entertain the idea that our results could be useful for the issue of taking the flat-space limit of holography. The fate of the UV boundary and Poincar\'e horizon in taking the flat space limit is particularly clear in our treatment, as is the emergence of the $d+1$ dimensional Poincar\'e group as an \.In\"on\"u-Wigner contraction of either $\mathfrak{so}(2,d)$ or $\mathfrak{so}(1,d+1)$. Here we content ourselves with remarking that the dilatation operator acting on bulk primary fields at $\epsilon = 0$ becomes the boost generator, $\bar{J}_{01}$ defined in appendix \ref{app:mink} i.e. the Rindler Hamiltonian. Primary states at $\epsilon = 0$ are constrained by \eqref{kill} to have vanishing null momentum, $\bar{P}_0 + \bar{P}_1$.

\section*{Acknowledgements}
It is a pleasure to acknowledge Alexandre Belin, Hong Liu, Jo\~ao Penedones, Riccardo Rattazzi, Arunabha Saha and Matthew Walters for discussions.
This work has been supported by the Fonds National Suisse de la Recherche Scientifique (FNS) through Project Grant 200021 162796 as well as the NCCR 51NF40-141869 ``The Mathematics of Physics" (SwissMAP).

\appendix
\section{The parameter $\epsilon$ and compactifications \label{app:theta}}
The goal for the metric \eqref{met} is to have the maximally symmetric Lorentzian spacetime which changes from flat-sliced AdS$_{d+1}$ to $\mathbb{R}_{1,d}$ to dS$_{d+1}$ upon variation of a parameter. To do this we introduced the vector $b$, and the coordinate $y^0 = \eta_{\mu\nu}b^\mu x^\nu$ which controls the distance from the boundary at $y^0=0$. A natural starting point is to consider the boundary placed at an angle $\theta$ in the $(x^1,x^0)$ plane. Then the vector $b$, together with an appropriate tangent vector $k$, may be defined as follows,
\bea
b &=& \cos\theta \frac{\partial}{\partial x^0} + \sin\theta \frac{\partial}{\partial x^1}\\
k &=& \sin\theta \frac{\partial}{\partial x^0} - \cos\theta \frac{\partial}{\partial x^1},
\eea
with norms,
\be
\eta_{\mu\nu}b^{\mu}b^{\nu} = -\cos(2\theta),\quad \eta_{\mu\nu}k^{\mu}k^{\nu}  = \cos(2\theta) \quad \eta_{\mu\nu}k^{\mu}b^{\nu}  = -\sin(2\theta).
\ee
Finally it is convenient to define, $\epsilon \equiv \cos(2 \theta)$, leading to the parameterisation by $\epsilon$ stated in the text. By varying $-1\leq \epsilon \leq 1$ we are varying the angle of the boundary in the plane $(x^1, x^0)$, varying it from timelike at $\theta = \pi/2$ $(\epsilon = -1)$ to null at $\theta=3\pi/4$ $(\epsilon = 0)$ to spacelike at $\theta = \pi$ $(\epsilon = 1)$.

Figure \ref{fig.slicings} shows compactified representations of the spacetime, with light rays travelling at 45 degrees, for various values of $\epsilon$. To construct these diagrams, we first factor out the conformal factor $1/(b\cdot x)^2$ from the line element \eqref{met}, leaving the expression for Minkowski space in $x^\mu$ coordinates. From here we use a standard compactification of the $x^0, x^1$ directions, 
\be
\tan(u\pm v) = x^1 \pm x^0.
\ee
Figure \ref{fig.slicings} shows lines of constant $y^1 \equiv \eta_{\mu\nu} k^\mu x^\nu$ and $y^0 \equiv \eta_{\mu\nu} b^\mu x^\nu$ in the $(u,v)$ plane for values $y^0 \geq 0$. The angle made by the surface $y^0=0$ at the origin of the $(u,v)$ plane is the same as that in the $(x^1,x^0)$ plane, i.e. it is given by $\theta$.  Negative values of $y^0$ cover the complementary region in the diamond defined by the points $(u,v) = (\pi/2,0), (0,\pi/2),(-\pi/2,0),(0,-\pi/2)$.

\section{Killing vectors from the embedding space\label{app:embed}}
An embedding space picture is convenient to access the isometries of the spacetime. The spacetimes can be represented as a hypersurface
\be
h_{AB} X^A X^B = \frac{L^2}{\epsilon}, 
\ee
where we have introduced the embedding space to have metric $h_{AB} = \text{diag}(-1,\epsilon,1,\ldots 1)$ where $A,B = 0,\ldots, d+1$. In the coordinate system used, $y^\mu$, the metric \eqref{gdef} can be reached using the following parameterisation of the hypersurface,
\bea
X^0 &=& \frac{L}{y^0}\;\frac{1+ \frac{(y^0)^2}{L^2}y\cdot y}{2}\nonumber\\
X^1 &=& \frac{L}{y^0}\;\left(y^1+ y^0\frac{\sqrt{1-\epsilon^2}}{\epsilon}\right)\nonumber\\
X^i  &=& \frac{L}{y^0}\; y^i\nonumber\\
X^{d+1} &=&  \frac{L}{y^0}\; \frac{1- \frac{(y^0)^2}{L^2}y\cdot y}{2}.\label{bigX}
\eea
For these choices we recover \eqref{gdef}, i.e.,
\be
ds^2 = h_{AB} dX^A dX^B =  g_{\mu\nu}dy^\mu dy^\nu.
\ee
The isometries of the $d+1$ dimensional spacetime of interest are then inherited from the isometries of $h_{AB}$, the Killing vectors $L_{AB} = X_A \partial_B - X_B \partial_A$. In particular, the vectors given in \eqref{Killing} are given by,
\bea
D  &=& -i L_{0,d+1}\nonumber\\
P_a &=& i (L_{0,a}-L_{d+1,a})\nonumber\\
J_{ab} &=& -i L_{a,b}\nonumber\\
K_a &=& i (L_{0,a}+L_{d+1,a}).
\eea

\section{The algebra at $\epsilon = 0$\label{app:mink}}
For $\epsilon = 0$ the maximally symmetric spacetime described by the line element \eqref{met} may be written in a coordinate system $\bar{x}^\mu$ such that the line element is $ds^2 = \eta_{\mu\nu} d\bar{x}^\mu d\bar{x}^\nu$. This familiar presentation of Minkowski space has the following isometries corresponding to $d+1$ dimensional translations, rotations and boosts
\bea
\bar{P}_\mu &=& -i \frac{\partial}{\partial\bar{x}^\mu}\\
\bar{J}_{\mu\nu} &=& -i\left(\bar{x}_\mu\frac{\partial}{\partial\bar{x}^\nu} - \bar{x}_\nu\frac{\partial}{\partial\bar{x}^\mu}\right)
\eea
obeying the Poincar\'e algebra in $d+1$ dimensions,
\bea
\left[\bar{J}_{\mu\nu},\bar{J}_{\rho \sigma}\right] &=& i \left(\eta_{\mu \rho}\bar{J}_{\nu \sigma} -\eta_{\mu \sigma}\bar{J}_{\nu \rho}+\eta_{\nu \sigma}\bar{J}_{\mu \rho}-\eta_{\nu \rho}\bar{J}_{\mu \sigma}\right),\\
\left[\bar{J}_{\mu\nu}, \bar{P}_\rho\right] &=& i \left(\eta_{\rho \mu} \bar{P}_\nu - \eta_{\rho \nu} \bar{P}_\mu \right).
\eea
We can relate these generators to those considered in the main text ($D,P_a,K_a,J_{ab}$) i.e. \eqref{Killing} at $\epsilon = 0$ by first enacting the coordinate transformation to reach the form of the metric \eqref{met},
\bea
\bar{x}^\mu &=&\frac{L x^\mu - L^2 k^\mu+\left(\tfrac{1}{2}x^2 - L k\cdot x\right)b^\mu }{b\cdot x} \label{xbarmap}
\eea
and then composing with the transformation to the $y^\mu$ coordinates to reach the form of the metric \eqref{gdef},
\bea
\bar{x}^0 &=& \frac{L y^0-L^2 - \frac{(y^0)^2}{2L^2}y\cdot y}{\sqrt{2} y^0}\\
\bar{x}^1 &=& \frac{Ly^0-L^2 + \frac{(y^0)^2}{2L^2}y\cdot y}{\sqrt{2} y^0}\\
\bar{x}^i &=& \frac{Ly^i}{y^0},
\eea
where we remind the reader that $i=2,\ldots d$. 
Using this coordinate mapping we can relate the generators in the text to the familiar generators and presentation of the Poincar\'e algebra above,
\bea
D &=& \frac{L}{\sqrt{2}}\left(\bar{P}_0 + \bar{P}_1\right) + \bar{J}_{01}\\
P_1 &=& \frac{1}{\sqrt{2}}\left(\bar{P}_1-\bar{P}_0\right)\\
P_i &=& \bar{P}_i + \frac{1}{\sqrt{2}L}\left(\bar{J}_{0i}-\bar{J}_{1i}\right)\\
J_{1i} &=& L \bar{P}_i\\
J_{ij} &=& \bar{J}_{ij}\\
K_1 &=& -\sqrt{2}L^2 \left(\bar{P}_0+\bar{P}_1\right)\\
K_i &=& \sqrt{2}L\left(\bar{J}_{0i}+\bar{J}_{1i}\right).
\eea
In this language, the quadratic Casimir \eqref{cas} is simply given by
\be
C_2 = \frac{L^2}{\epsilon}\left(\eta^{\mu\nu} \bar{P}_\mu \bar{P}_\nu\right).
\ee

\bibliographystyle{utphys}
\bibliography{gravitons}{}

\end{document}